\documentclass[twocolumn,showpacs,preprintnumbers,amsmath,amssymb]{revtex4}

\usepackage{graphicx}
\usepackage{dcolumn}
\usepackage{bm}

\begin{document}

\title{Slower Speed and Stronger Coupling:\\
Adaptive Mechanisms of Self-Organized Chaos Synchronization}

\author{Xiao Fan Wang}
 \email{X.Wang@bristol.ac.uk}
\affiliation{Department of Mechanical Engineering\\
University of Bristol, Bristol, BS8 1TR, UK}

\date{\today}

\begin{abstract}
We show that two initially weakly coupled chaotic systems can achieve self-organized 
synchronization by adaptively reducing their speed and/or enhancing the coupling
strength. Explicit adaptive algorithms for speed-reduction and
coupling-enhancement are provided. We apply these algorithms to the self-organized
synchronization of two coupled Lorenz systems. It is found that after a long-time self-organized process,
the two coupled chaotic systems can achieve synchronization with almost minimum required coupling-speed ratio.

\end{abstract}

\pacs{05.45.+b}

\maketitle

The ability of coupled nonlinear oscillators to synchronize with each other is a basis for the explanation of many
processes in nature.
 Traditionally, synchronization has been understood as the ability
of coupled periodic oscillators with different frequencies to switch their behavior from the regime of
independent oscillations to the regime of cooperative oscillations, as the strength of the coupling is
increased. In recent years, 
synchronization in coupled chaotic systems has become an important researcch field
with applications in many areas of science and technology [1, 2],
such as communications [3, 4], lasers [5], chemical and biological systems [6, 7].
It has been shown that two coupled identical chaotic systems
can change their behavior from uncorrelated chaotic oscillations to identical chaotic oscillations,
as the coupling strength is increased [1].
However, this result does not tell us how the coupling strength is increased as
two initially weakly coupled systems evolve into synchronization.

Recently, N\'eda {\sl et al.} proposed a new mechanism for self-organized synchronization 
in coupled periodic oscillators [8].
They investigated the development and dynamics of the rhythmic applause in
concert halls. They found that the mechanism lying at the heart of the synchronization process
is the period doubling of the clapping rhythm which leads to slower clapping modes. 

In this Letter, we propose a unified mechanism for self-organized synchronization of initially
weakly coupled chaotic systems. We show that two coupled systems can achieve self-organized 
synchronization by adaptively reducing their speed and/or enhancing the coupling
strength. We also provide explicit adaptive algorithms for speed-reduction and
coupling-enhancement.

Consider two linearly coupled identical chaotic systems described by
\begin{equation}
\dot{\bf x}(t)=\tau (t){\bf F}({\bf x}(t))+{\bf K}(t)({\bf y}(t)-{\bf x}(t)),
\end{equation}
\begin{equation}
\dot{\bf y}(t)=\tau (t){\bf F}({\bf y}(t))+{\bf K}(t)({\bf x}(t)-{\bf y}(t)),
\end{equation}
where ${\bf x}=[x_1, x_2, \cdots, x_n]^T, {\bf y}=[y_1, y_2, \cdots, y_n]^T \in \Re^n$ are the states of the coupled systems. 
$\tau (t)>0$ is called the {\sl speed factor} of the coupled systems. The coupling matrix
${\bf K}(t)=diag(k_1(t), k_2(t), \cdots, k_n(t))$ is assumed to be a diagonal matrix with $k_i(t)\equiv k(t)>0$
for a particular $i$ and $k_j(t)\equiv 0$ for $j\neq i$. This means that systems (1) and (2) are linearly coupled
through their $i$th state variables, and $k(t)$ represents the {\sl coupling strength}. 
We can rewrite systems (1) and (2) as follows:
\begin{equation}
\dot{\bf x}(t)=\tau (t) \{{\bf F}({\bf x}(t))+\frac{1}{\tau (t)}{{\bf K}(t)}({\bf y}(t)-{\bf x}(t))\},
\end{equation}
\begin{equation}
\dot{\bf y}(t)=\tau (t) \{{\bf F}({\bf y}(t))+\frac{1}{\tau (t)}{{\bf K}(t)}({\bf x}(t)-{\bf y}(t))\}.
\end{equation}

Note that if $\tau(t) \equiv \tau_0>0$ and $k(t) \equiv k_0>0$ are two constants, and the $(n-1)$-dimensional
subsystem
\begin{equation}
\dot{x}_j(t)=f_j({\bf x}(t)), \quad j=1, 2, \cdots, i-1, i+1, \cdots, n,
\end{equation}
is asymptotically stable [2], then there exists a critical value $\bar k>0$ such that
the two coupled systems (3) and (4) will synchronize in the sense that 
${\bf x}(t)-{\bf y}(t) \to \bf 0$ as $t \to \infty$ when $k_0/\tau_0>\bar k$. 
The dynamics of the synchronization states 
${\bf x}(t)={\bf y}(t)$ is governed by
\begin{equation}
\dot{\bf x}(t)=\tau_0 {\bf F}({\bf x}(t)),
\end{equation}
which has the same attractor for different nonzero values of $\tau_0$. However, 
the value of $\tau_0$ determines the varying speed of the synchronization state. 
The lower the value of $\tau_0$, the lower the state varying speed.
For example, let $\bf F$ be the vector of the chaotic Lorenz system [9]. Then (6) becomes
\begin{equation}
\left(
\begin{array}{c}
\dot x_1(t)\\
\dot x_2(t)\\
\dot x_3(t)
\end{array}\right)
=\tau_0
\left(
\begin{array}{c}
\sigma (x_2-x_1)\\
rx_1-x_2-x_1x_3\\
x_1x_2-bx_3
\end{array}\right)\;,
\end{equation}
where we take $\sigma=10$, $b=8/3$ and $r=28$. For $\tau_0=1,\; 0.5,\; 0.2$, system (7)
has the same chaotic attractor but different state varying speed, as shown in Fig. 1.
\begin{figure}
\includegraphics{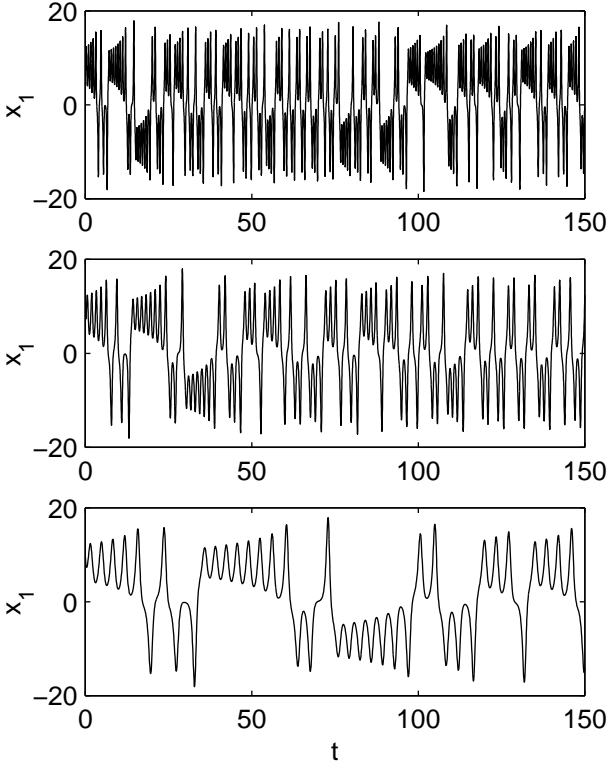}
\caption{\label{fig:epsart} Trajectories of $x_1(t)$ of system (7) with $\tau=1$ (upper), 
$0.5$ (middle) and $0.2$ (bottom), respectively.}
\end{figure}
\begin{figure}
\includegraphics{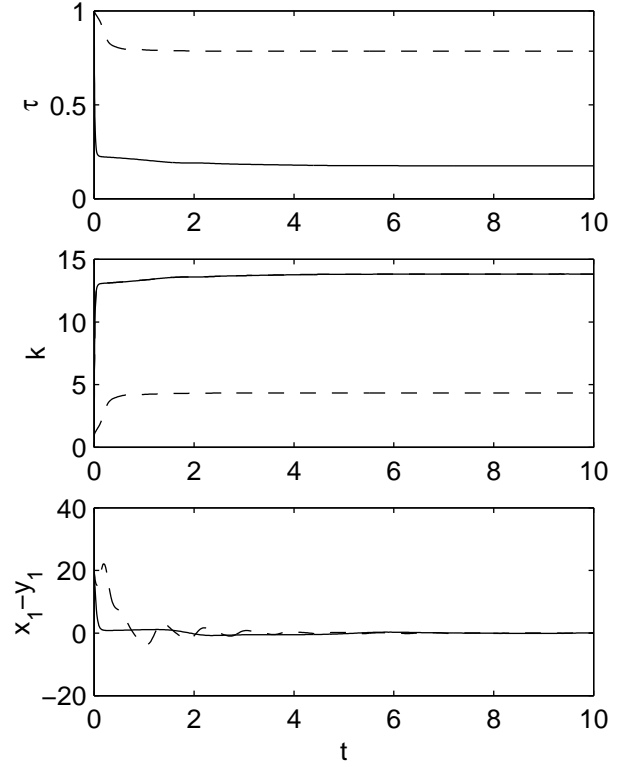}
\caption{\label{fig:epsart} Adaptive synchronization of the coupled Lorenz systems (13)-(16)
with $\gamma=0.2$ (full line) and $0.002$ (dashed line).}
\end{figure}
\begin{figure}
\includegraphics{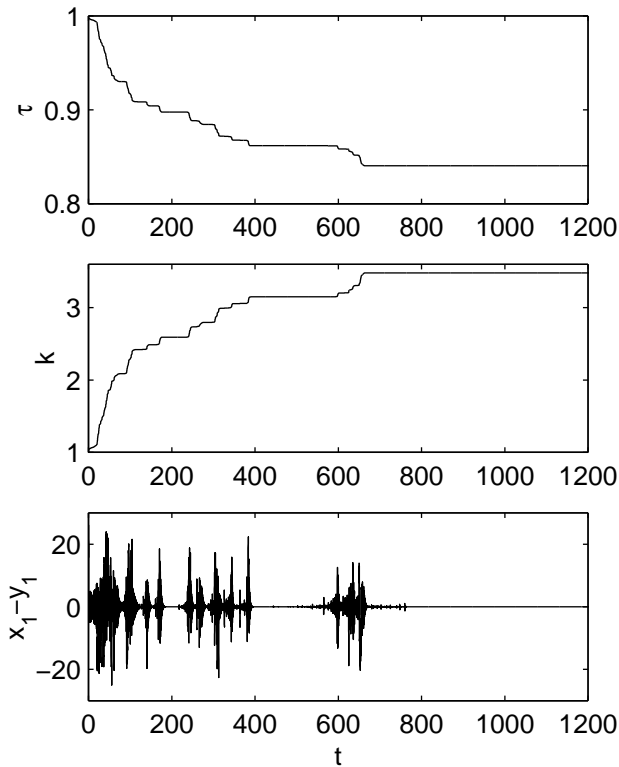}
\caption{\label{fig:epsart} Adaptive synchronization of the coupled Lorenz systems (13)-(16)
with $\gamma=0.00002$.}
\end{figure}
\begin{figure}
\includegraphics{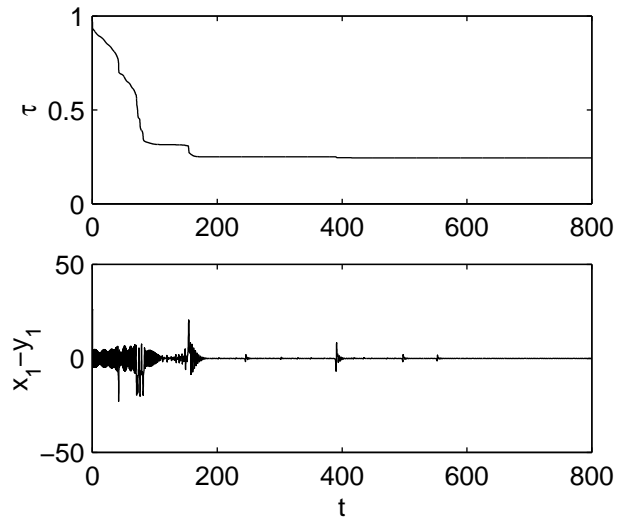}
\caption{\label{fig:epsart} Synchronization of the coupled Lorenz systems (13) and (14)
via adaptive speed-reduction algorithm (15) with gain $\gamma_\tau=0.0005$.}
\end{figure}
\begin{figure}
\includegraphics{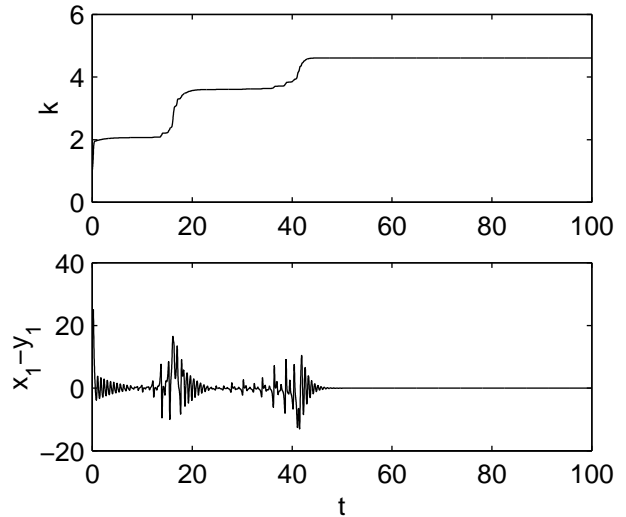}
\caption{\label{fig:epsart} Synchronization of the coupled Lorenz systems
via adaptive coupling-enhancement algorithm (16) with gain $\gamma_k=0.0005$.}
\end{figure}

One may view the coupling-speed ratio $k/\tau$ as the cost of the coupled systems and 
the critical value $\bar k$ as the minimum cost required to achieve synchronization.
If the value of $\bar k$ is known, then one may take a smaller $\tau_0$ and/or a larger $k_0$
so as to guarantee $k_0/\tau_0>\bar k$ .
However, to find the critical value $\bar k$ at the beginning, the complete model of the coupled systems and perhaps
the exact initial conditions must be known. Thus it is interesting to see how two initially weakly coupled systems can
adaptively update their varying speed and coupling strength according to measured synchronization
error until synchronization occurs.

Without loss of generality, we assume $\tau_0=1$ and $k_0<\bar k$.
The adaptive algorithms for the update of speed factor and coupling strength are described by
\begin{equation}
\dot{\tau}(t)=\gamma_\tau (x_i(t)-y_i(t))^2 (\tau^*-\tau (t))\;, \quad \tau (t)=\tau_0,
\end{equation}
\begin{equation}
\dot k(t)=\gamma_k (x_i(t)-y_i(t))^2(k^*-k(t))\;, \quad k(0)=k_0\;,
\end{equation}
where $\gamma_\tau$ and $\gamma_k$ are positive adaptive gains. $\tau^*>0$ and $k^*>0$
represents the minimum allowed speed factor and maximum allowed coupling strength, respectively.
Therefore, $k^*/\tau^*$ represents the maximum allowed coupling-speed ratio (cost) of the coupled systems. A necessary
 condition for synchronization is $k^*/\tau^*>\bar k$. 

From Eq. (8), $\tau (t)$ is a decreasing function and $\tau (t) \geq \tau^*$;
From Eq. (9), $k(t)$ is an increasing function and $k(t) \leq k^*$. 
Therefore, there exist two constants $\hat \tau$ and $\hat k$ such that
\begin{equation}
\lim_{t\rightarrow \infty }\tau (t)=\hat \tau\;,\quad \tau^*\leq\hat\tau\leq\tau_0\;,
\end{equation}
\begin{equation}
\lim_{t\rightarrow \infty }k(t)=\hat k\;,\quad k_0\leq \hat k\leq k^*\;,
\end{equation}

If $\hat \tau=\tau^*$ and $\hat k=k^*$, then $\hat k/\hat \tau=k^*/\tau^*>\bar k$, which implies
that the coupled systems will achieve asymptotic synchronization. On the other hand, if
$\hat \tau \neq \tau^*$ and/or $\hat k\neq k^*$, then from (8) and (9), we must have
\begin{equation}
\lim_{t\rightarrow \infty }(x_i(t)-y_i(t))=0,
\end{equation}
which, together with the stability of the subsystem (5), imply that the coupled systems (1)
and (2) will also achieve asymptotic synchronization. 

As an example, we consider two coupled Lorenz systems described by
\begin{equation}
\left(
\begin{array}{c}
\dot x_1\\
\dot x_2\\
\dot x_3
\end{array}\right)
=\tau (t)
\left(
\begin{array}{c}
\sigma (x_2-x_1)+k(t)(y_1-x_1)\\
rx_1-x_2-x_1x_3\\
x_1x_2-bx_3
\end{array}\right)\;,
\end{equation}
\begin{equation}
\left(
\begin{array}{c}
\dot y_1\\
\dot y_2\\
\dot y_3
\end{array}\right)
=\tau (t)
\left(
\begin{array}{c}
\sigma (y_2-y_1)+k(t)(x_1-y_1)\\
ry_1-y_2-y_1y_3\\
y_1y_2-by_3
\end{array}\right)\;,
\end{equation}
with the following two adaptive algorithms:
\begin{equation}
\dot \tau(t)=\gamma_\tau (x_1-y_1)^2(\tau^*-\tau(t))\;, 
\end{equation}
\begin{equation}
\dot k(t)=\gamma_k (x_1-y_1)^2(k^*-k(t))\;.
\end{equation}

In the following simulations, the initial states are taken as
 ${\bf x}(0)=[10,\; 2,\; 20]^T$ and ${\bf y}(0)=[-10,\; -2,\; 0]^T$. 
We find through simulation that the critical value is $\bar k\approx 3.950$. Other parameters were chosen
as follows: $\tau_0=1$, $k_0=1$, $\tau^*=0.1$,  and $k^*=15$. 

We take $\gamma_\tau=\gamma_k=\gamma$. Fig. 2 and 3 show the simulation results for $\gamma=0.2$,
$0.002$ and $0.00002$.
For large adaptive gain $\gamma=0.2$, $\tau (t)$ decreases fast to $\hat \tau\approx 0.17$ which is close to
the minimum allowed value $\tau^*$ and $k(t)$ increases fast to $\hat k\approx13.82$ which is close to 
the maximum allowed value $k^*$ (Fig. 2).
In this case, although the time to achieve synchronization is quite short (Fig. 2), the cost 
$\hat k/\hat \tau\approx 81.29$ is much higher than the minimum required cost $\bar k$.
On the other hand, for sufficiently small adaptive gain $\gamma=0.00002$,
 $\tau (t)$ decreases to $\hat \tau=0.84$ and $k(t)$ increases to $\hat k=3.48$ very slowly, and therefore,
 the time to achieve synchronization is quite long (Fig. 3); However, in this case, 
we find that $\hat k/\hat\tau\approx4.14$ which is very close to the critical value $\bar k$. 
This implies that after a long-time self-organized process, the coupled systems can achieve synchronization
with almost minimum required cost.
For middle adaptive gain
 $\gamma=0.002$, $\tau (t)$ decreases fast to $\hat \tau=0.78$ and 
$k(t)$ increases fast to $\hat k=4.34$. In this case, the coupled systems can achieve fast synchronization with low cost 
 $\hat k/\hat\tau\approx5.52$ (Fig. 2).

Now suppose that the coupling strength can not be changed directly,
i.e., $k(t)\equiv k_0$. We show that the coupled systems (13) and (14) can achieve 
synchronization by adaptively tunning their speed according to algorithm (15) alone. In this case, 
the critical speed factor is $\bar \tau=k_0/\bar k \approx 0.253$. We find that
a large adaptive gain would lead to fast synchronization, but at the same time, 
$\hat \tau$ is very close to the minimum allowed value $\tau^*$ which implies that
the varying speed of the synchronization state might be too slow to be efficient.
For sufficiently small adaptive gain $\gamma_\tau$, the time to achieve synchronization 
may be quite long, but $\hat \tau$ is very close to the critical value $\bar \tau$ which
implies that after a long-time self-organized process, the coupled systems can
achieve synchronization and keep the varying speed of the synchronization state as fast as possible.
Fig. 4 shows the synchronization process for $\gamma_\tau=0.0005$.

If the coupled systems (13) and (14) wish to keep their speed unchanged, i.e., $\tau(t)\equiv \tau_0=1$, 
then they can achieve synchronization by adaptively increase the coupling strength according to 
algorithm (16) alone. In this case, the critical coupling strength is $\bar k\approx 3.950$.
A sufficiently large adaptive gain would lead to fast synchronization, but
$\hat k$ is very close to the maximum allowed value $k^*$ which 
implies that the coupling strength might be too strong to be safe.
For sufficiently small adaptive gain $\gamma_k$, the time to achieve synchronization
 may be quite long, and $\hat k$ is very close to the critical value $\bar k$ which  
implies that after a long-time self-organized process, the coupled systems can
achieve synchronization and keep the coupling strength as small as possible.
Fig. 5 shows the synchronization process for $\gamma_k=0.0005$. 

In summary, we proposed adaptive speed-reduction and/or coupling enhancement algorithms 
for self-organized synchronization in two initially weakly coupled chaotic systems. We showed that
if the adaptive gains are sufficiently small, then after a long-time self-organized process,
the two coupled chaotic systems can achieve synchronization with almost minimum required coupling-speed ratio.

\newpage 

\noindent{\Large\bf References}
\begin{itemize}
\item[[1]]
 H. Fujisaka and T. Yamada, Prog.\ Theor.\ Phys. {\bf 69}, 32 (1984).
\vspace{-.5em}
\item[[2]]
L. M. Pecora and T. L. Carroll, Phys.\ Rev.\ Lett. {\bf 64}, 821(1990).
\vspace{-.5em}
\item[[3]]
K. M. Cuomo and A. V. Oppenheim, Phys.\ Rev.\ Lett. {\bf 71}, 65(1993).
\vspace{-.5em}
\item[[4]]
L. Kocarev and U. Parlitz, Phys.\ Rev.\ Lett. {\bf 74}, 5028(1995).
\vspace{-.5em}
\item[[5]]
R. Roy and K. S. Thornburg, Phys.\ Rev.\ Lett. {\bf 72}, 2009(1994).
\vspace{-.5em}
\item[[6]]
S. K. Han, C. Kurrer and Y. Kuramoto, Phys.\ Rev.\ Lett. {\bf 75}, 3190(1995).
\vspace{-.5em}
\item[[7]]
B. Blasius, A. Huppert and L. Stone, Nature {\bf 399}, 354 (1999).
\vspace{-.5em}
\item[[8]]
Z. N{\'e}da, E. Ravasz, Y. Brechet, T. Vicsek and A.-L. Barab{\'a}si, Nature,
{\bf 403}, 849 (2000).
\vspace{-.5em}
\item[[9]]
J. Gukenheimer and P. Holmes, {\sl Nonlinear Oscillations, Dynamical Systems, 
and Bifurcations of Vector Fields} (Springer-Verlag, New York, 1983).
\vspace{-.5em}
 
\end{itemize}

\end{document}